\documentclass[preprint,12pt]{elsarticle}

\usepackage{amssymb,amsmath}

\usepackage{color}
\newcommand{\vicente}[1]{{ #1}}

\newcommand\beq{\begin{equation}}
\newcommand\eeq{\end{equation}}
\newcommand\beqa{\begin{eqnarray}}
\newcommand\eeqa{\end{eqnarray}}
\newcommand{\dd}{\text{d}}

\journal{Physica A}

\usepackage{graphicx}% Include figure files

\begin{document}

\begin{frontmatter}

%% Title, authors and addresses

%% use the tnoteref command within \title for footnotes;
%% use the tnotetext command for theassociated footnote;
%% use the fnref command within \author or \address for footnotes;
%% use the fntext command for theassociated footnote;
%% use the corref command within \author for corresponding author footnotes;
%% use the cortext command for theassociated footnote;
%% use the ead command for the email address,
%% and the form \ead[url] for the home page:
%% \title{Title\tnoteref{label1}}
%% \tnotetext[label1]{}
%% \author{Name\corref{cor1}\fnref{label2}}
%% \ead{email address}
%% \ead[url]{home page}
%% \fntext[label2]{}
%% \cortext[cor1]{}
%% \address{Address\fnref{label3}}
%% \fntext[label3]{}

\title{Influence of a \vicente{drag} force on linear transport in low-density gases. Stability analysis}
\author{Jos\'e Carlos P\'erez-Fuentes}
\author{Vicente Garz\'o\corref{cor1}}
\ead{vicenteg@unex.es}
\ead[url]{http://www.unex.es/eweb/fisteor/vicente/}
%\fntext[label2]{}
\cortext[cor1]{Corresponding author. Tel.: +34 924289527.}
%\email{vicenteg@unex.es} \homepage{http://www.unex.es/eweb/fisteor/vicente/}
%\author{Mois\'es G. Chamorro}
%\author{Francisco Vega Reyes}
%\email{fvega@unex.es} \homepage{http://www.unex.es/eweb/fisteor/fran/}
\address{Departamento de F\'{\i}sica, Universidad de Extremadura, E-06071 Badajoz, Spain}

%\title{}

%% use optional labels to link authors explicitly to addresses:
%% \author[label1,label2]{}
%% \address[label1]{}
%% \address[label2]{}

%\author{}

%\address{}

\begin{abstract}

The transport coefficients of a dilute classical gas in the presence of a \vicente{drag} force proportional to the velocity of the particle are determined from the Boltzmann equation. \vicente{The viscous drag force could model the friction of solid particles with a surrounding fluid (interstitial gas phase).} First, when the \vicente{drag} force is the only external action on the state of the system, the Boltzmann equation admits a Maxwellian solution $f_0(\mathbf{v},t)$ with a time-dependent temperature. Then, the Boltzmann equation is solved by means of the Chapman-Enskog expansion around the local version of the distribution $f_0$ to obtain the relevant transport coefficients of the system: the shear viscosity $\eta$, the thermal conductivity $\kappa$, and a new transport coefficient $\mu$ (which is also present in granular gases) relating the heat flux with the density gradient. \vicente{The results indicate that while $\eta$ is not affected by the drag force, the impact of this force on the transport coefficients $\kappa$ and $\mu$ may be significant}.  Finally, a stability analysis of the linear hydrodynamic equations with respect to the time-dependent equilibrium state is performed, showing that the onset of instability is associated with the transversal shear mode that could be unstable for wave numbers smaller than a certain critical wave number.

\end{abstract}

\begin{keyword}

Boltzmann kinetic equation \sep \vicente{drag force} \sep transport coefficients \sep stability analysis.

\PACS 05.20.Dd \sep 45.70.Mg \sep 51.10.+y \sep 05.60.-k

%\pacs{}

%% keywords here, in the form: keyword \sep keyword

%% PACS codes here, in the form: \PACS code \sep code

%% MSC codes here, in the form: \MSC code \sep code
%% or \MSC[2008] code \sep code (2000 is the default)

\end{keyword}

\end{frontmatter}

%\linenumbers

%\draft

%\date{\today}
%\maketitle

\section{Introduction}
\label{sec1}

It is quite usual in computer simulations to control the temperature of a system by the action of a velocity-dependent external force. This type of external forces are usually called ``thermostats'' \cite{EM84,EH86,EM90}. In the case of ordinary fluids, the possibility of controlling the temperature is quite important from a practical point of view since it allows for instance to compensate for dissipative heating effects and achieve a non-equilibrium \emph{steady} state. In the case of the so-called granular gases (namely, a gas constituted by hard spheres that collide inelastically) \cite{G03,BP04}, the presence of an external driving force allows to compensate for the energy dissipated by collisions and maintain the system in rapid flow conditions \cite{simulations}. Nevertheless, in spite of its practical importance, the understanding of the influence of these forces on the transport properties of the system is not completely understood yet \cite{DSBR86,GS03}.

One of the simplest choices for the external force $\mathbf{F}$ seems to be a homogeneous force proportional to the velocity of the particle $\mathbf{v}$, \vicente{namely, $\mathbf{F}(\mathbf{v})=m\gamma \mathbf{v}$, where $m$ is the mass of a particle and $\gamma$ is a drag or friction coefficient}. This drag force (whose form is based on Gauss' principle of least constraint \cite{EM90}) has been mainly used in non-equilibrium molecular dynamics simulations of shear flows to prevent the viscous heating effect and measure the rheological properties of the fluid (such as the non-Newtonian shear viscosity) in steady state conditions.
\vicente{In this case, the friction coefficient $\gamma$ is coupled to the shear rate so that the drag force acts as a thermostatic force to keep the temperature constant}. A natural question is to gauge the relationship between the transport properties of the system obtained in the presence of or in the absence of the thermostat. Thus, in the case of the Boltzmann equation for a dilute gas of Maxwell molecules under uniform shear flow, there is a close relationship between the velocity distribution function with and without the thermostat \cite{DSBR86}. On the other hand, for non-Maxwell molecules, the above relationship does not exist and the impact of the force on transport properties can be significant \cite{GS03}. In particular, the non-Newtonian shear viscosity of an unforced ordinary gas differs from the one obtained when the viscous heating is controlled.

\vicente{Apart from the use of nonconservative forces as thermostats, the viscous drag force could also model the influence of gas phase on the dynamic properties of solid particles in gas-solid flows \cite{K90,KH01,H13,WGZS14}. In fact, the Boltzmann kinetic equation employed here (see Eq.\ \eqref{3.2} below) to determine the transport coefficients reduces (in the case of elastic collisions and when the mean flow velocities of solid and gas phases coincide) to a recent kinetic equation \cite{GTSH12} proposed to analyze dynamic properties of monodisperse gas-solid suspensions. In this context, we expect that our study may have applications in mesoscopic systems, such as colloids and suspensions \cite{K90,KH01,J00,BGP11}.}

To the best of our knowledge, the only attempt to analyze the influence of \vicente{the drag force} on transport was carried out years ago \cite{GSB90} for the self-diffusion coefficient. Two different physical situations were considered: (i) when the system as a whole is at equilibrium and (ii) when the system is under uniform shear flow. \vicente{While in the first case the friction coefficient $\gamma$ is a free parameter independent of the spatial gradients, the parameter $\gamma$ is a shear-rate dependent function in the shear flow case}. The results clearly show that in general the expression of the self-diffusion coefficient \vicente{at equilibrium} is strongly modified with respect to the conventional (unforced) situation. The aim of this paper is to extend the calculations carried out in Ref.\ \cite{GSB90} for the self-diffusion coefficient to the remaining transport coefficients of the system. \vicente{The fact that the drag force mimics the effect of the interstitial fluid on solid particles in gas-solid suspensions justifies the physical interest of the present study}. In this paper, the transport coefficients of the system are obtained by solving the Boltzmann equation by means of the Chapman-Enskog method \cite{CC70} to first order in the spatial gradients. As noted in Ref.\ \cite{GSB90}, since the main effect of the external force is just to change the temperature in time, the zeroth-order approximation $f^{(0)}$ in the Chapman-Enskog expansion is a (local) time dependent Maxwell-Boltzmann distribution function. This fact introduces additional technical difficulties not present in the absence of the drag force. As expected, while all the transport coefficients can be \emph{exactly} obtained from the Boltzmann equation for Maxwell molecules, one has to resort to a simple kinetic model equation to determine them for non-Maxwell molecules.  \vicente{When the friction coefficient $\gamma$ is not uniform, there is a \emph{new} contribution (not present in the unforced case) to the heat flux proportional to the density gradient.} This contribution defines a transport coefficient $\mu$ which is also present in the case of granular gases \cite{BDKS98,GD99}. On the other hand, the origin of the coefficient $\mu$ here is associated with the presence of the viscous drag force ($\gamma \neq 0$) rather than the inelastic character of collisions. Note that the presence of this extra term in the heat current could give rise to an anomalous heat flow since the flow of heat from cold to hot could be permitted \cite{CW07}.

The knowledge of the transport coefficients, allows us to perform a linear stability analysis of the hydrodynamic equations with respect to the time-dependent equilibrium state in order to identify the conditions for stability as functions of the wave vector and the friction coefficient $\gamma$. As we shall see, linear stability shows two transversal (``shear'') modes and a longitudinal (``heat'') mode to be unstable with respect to long enough wavelength excitations. This kind of instability follows essentially from the presence of a sink term in the equation for the balance of energy \vicente{and presents some similarities with the clustering instability for granular gases detected in the seminal papers of Golhirsch and Zanetti \cite{GZ93} and McNamara and Young \cite{M93} in the so-called homogeneous cooling state.}

The plan of the paper is as follows. In Section \ref{sec2}, the drag force to be considered is defined. It is shown that whether this force is the only external action on the system, the system reaches in time an equilibrium-like state but with a time-dependent temperature. Then, the corresponding Boltzmann equation with the presence of the drag force is solved by means of the Chapman-Enskog expansion around this time-dependent equilibrium. The transport coefficients are explicitly determined in Section \ref{sec4} from the Boltzmann equation for Maxwell molecules and from the well-known Bhatnagar-Groos-Krook (BGK) model \cite{C75} for general repulsive potentials. Section \ref{sec5} is devoted to the linear stability analysis around the time-dependent equilibrium state while the paper is closed in Section \ref{sec6} with some conclusions.

\section{Boltzmann kinetic equation. Nonconservative forces}
\label{sec2}

We consider a dilute classical gas subjected to a nonconservative external force $\mathbf{F}(\mathbf{r},\mathbf{v},t)$. In the low-density regime, one can assume that there are no correlations between the velocities of two particles that are about to collide (molecular chaos hypothesis) and hence, the two-body distribution function factorizes into the product of the one-particle distribution functions $f(\mathbf{r},\mathbf{v},t)$. This distribution obeys the nonlinear Boltzmann equation
\begin{equation}
\label{2.1}
\frac{\partial f}{\partial t}+\mathbf{v}\cdot \nabla f+\frac{\partial}{\partial \mathbf{v}}\cdot \left(\frac{\mathbf{F}}{m}f\right)=J[\mathbf{v}|f,f],
\end{equation}
where $m$ is the mass of a particle and $J[f,f]$ is the Boltzmann collision operator \cite{C75}. At a kinetic level, the knowledge of the velocity distribution function $f(\mathbf{r},\mathbf{v},t)$ provides all the relevant information on the state of the system. In particular, the first few velocity moments of $f$ are the local number density
\beq
\label{2.2}
n({\bf r}, t)=\int \dd\mathbf{v}\;f({\bf r},{\bf v},t),
\eeq
the local flow velocity
\begin{equation}
\label{2.3}
{\bf U}({\bf r}, t)=\frac{1}{n({\bf r}, t)}\int
\dd{\bf v} \;{\bf v} f({\bf r},{\bf v},t),
\end{equation}
and the local temperature
\begin{equation}
\label{2.4}
T({\bf r}, t)=\frac{m}{3 n({\bf r}, t)}\int  \dd{\bf v} \;V^2({\bf r}, t) f({\bf r},{\bf v},t),
\end{equation}
where ${\bf V}({\bf r},t)\equiv {\bf v}-{\bf U}({\bf r}, t)$ is the peculiar velocity. The macroscopic balance equations for density, momentum, and energy follow directly from Eq.\  \eqref{2.1} by multiplying with $1$,
$m{\bf v}$, and $\frac{1}{2}mv^2$ and integrating over ${\bf v}$. The result is \cite{GS03}
\begin{equation}
\label{2.5} D_{t}n+n\nabla \cdot {\bf U}=0\;,
\end{equation}
\begin{equation}
\label{2.6} \rho D_{t}\mathbf{U}+\nabla \cdot \mathsf{P}=\boldsymbol{\sigma}_U,
\end{equation}
\begin{equation}
\label{2.7} \frac{3}{2}n k_B D_{t}T+\nabla \cdot {\bf q}+\mathbf{P}:\nabla \mathbf{U} =\sigma_T\;.
\end{equation}
Here, $D_{t}=\partial _{t}+{\bf U}\cdot \nabla$, $\rho=mn$ is the mass density, and the microscopic
expressions for the pressure tensor ${\sf P}$ and the heat flux ${\bf
q}$ are given, respectively, by
\begin{equation}
{\sf P}=\int \dd{\bf v}\,m{\bf V}{\bf V}\,f({\bf v}), \quad {\bf q}=\int \dd{\bf v}\,\frac{1}{2}m V^{2}{\bf V}\,f({\bf v}).
\label{2.8}
\end{equation}
%\begin{equation}
%{\bf q}=\int \dd{\bf v}\,\frac{1}{2}m V^{2}{\bf V}\,f({\bf v}). \label{2.9}
%\end{equation}
In addition, in Eqs.\ \eqref{2.6} and \eqref{2.7} we have introduced the production of momentum $\boldsymbol{\sigma}_U$ and energy $\sigma_T$ due to the external force. They are given by
\beq
\label{2.9}
\boldsymbol{\sigma}_U=\int \dd\mathbf{v}\; \mathbf{F} f({\bf v}), \quad \sigma_T=\int \dd\mathbf{v}\; \mathbf{V}\cdot \mathbf{F} f({\bf v}).
\eeq
%\beq
%\label{2.10}
%\sigma_T=\int \dd\mathbf{v}\; \mathbf{V}\cdot \mathbf{F} f({\bf v}).
%\eeq

\subsection{Homogeneous states}

In computer simulations \cite{EM90,simulations}, the external \emph{nonconservative} forces $\mathbf{F}$ have been widely employed to control the temperature of the system. The simplest possibility seems to be a homogeneous nonconservative force proportional to the velocity of the particle \cite{EM90}. In this case,
\beq
\label{2.11}
\mathbf{F}(\mathbf{v},t)=m\gamma (t) \mathbf{v},
\eeq
where the amplitude of the external force $\gamma$ can be a function of time. \vicente{In the case of sheared ordinary fluids, the friction coefficient $\gamma$ is a negative shear-rate dependent function chosen to compensate for the viscous heating produced by shear work \cite{GS03}. On the other hand, in the case of granular fluids (a gas whose particles collide inelastically), $\gamma$ is a positive coefficient (coupled to the coefficient of restitution of inelastic collisions) chosen to compensate for the energy lost by collisions \cite{MG02}. In both cases, the external force plays the role of an external thermostat introduced to maintain the system in stationary conditions. However, as said in the Introduction, the viscous drag force \eqref{2.11} has been also used to model the friction of solid particles with the surrounding fluid in gas-solid suspensions \cite{KH01,H13,WGZS14,GTSH12}. In this case and for classical gases, $\gamma$ is a \emph{negative} coefficient independent of the imposed spatial gradients. This will be the point of view adopted in this paper}.

In the homogeneous state, $n$ and $T$ are spatially uniform, and with an appropriate selection of the frame of reference, the mean flow velocity vanishes ($\mathbf{U}=\mathbf{0}$). Moreover, the production terms are $\boldsymbol{\sigma}_U=\mathbf{0}$ and $\sigma_T=3 n k_B T$ so that,
the only relevant balance equation is that of the temperature \eqref{2.7} which reads
\beq
\label{2.12}
\frac{\partial T}{\partial t}=2 \gamma T.
\eeq
Under these conditions, it is easy to see that the Boltzmann equation \eqref{2.1} admits the solution \cite{GSB90}
\beq
\label{2.13}
f_0(\mathbf{v},t)=n\left(\frac{m}{2\pi k_BT(t)}\right)^{3/2} \text{exp}\left(-\frac{mv^2}{2k_BT(t)}\right),
\eeq
where $T(t)$ is the solution to Eq.\ \eqref{2.12}. Thus, the system is in a ``time-dependent equilibrium'' state. The time evolution of the temperature depends on the choice of the friction coefficient $\gamma(t)$. In particular, the choice $\gamma \equiv \text{const.}$ leads to an exponential behavior. Moreover, in Ref.\ \cite{GSB90} it was also proved an $H$-theorem for the distribution $f_0$ in the sense that, starting from any initial condition and in the presence of the drag force \eqref{2.11}, the velocity distribution function $f(\mathbf{r},\mathbf{v},t)$ reaches in the long time limit the Maxwellian form \eqref{2.13} with a time-dependent temperature.

%Now, the objective is to analyze the effect of the nonconservative force \eqref{2.11} on the relevant transport coefficients of the system. This will be done in the next Section.

\section{Chapman-Enskog method}
\label{sec3}

The homogeneous state briefly described in Section \ref{sec2} can be disturbed by the presence of small spatial gradients. These gradients give rise to contributions to the momentum and heat fluxes which are characterized by the transport coefficients. The determination of these coefficients is one of the main goals of this paper.

Since we have to consider an inhomogeneous state, one has first to generalize the form of the drag force \eqref{2.11} to this situation. \vicente{Since the influence of the drag force on transport attempts to model the effect of gas phase into the dynamics of solid particles, here  we will assume that the external drag force is given by}
\beq
\label{3.1}
\mathbf{F}=m\gamma (\mathbf{v}-\mathbf{U}_g),
\eeq
where $\mathbf{U}_g$ is a known reference velocity of the system. \vicente{Therefore, $\mathbf{U}_g$ can be interpreted as the mean velocity of gas surrounding the solid particles}. Thus, the Boltzmann equation \eqref{2.1} with the form \eqref{3.1} for the external force becomes
\beq
\frac{\partial f}{\partial t}+\mathbf{v}\cdot \nabla f+\gamma \Delta \mathbf{U} \cdot
\frac{\partial}{\partial\mathbf{v}}f+\gamma
\frac{\partial}{\partial\mathbf{v}}\cdot \mathbf{V} f=J[f,f], \label{3.2}
\eeq
where $\Delta \mathbf{U}=\mathbf{U}-\mathbf{U}_\text{g}$. When $\mathbf{U}=\mathbf{U}_g$ and $\gamma<0$, the Boltzmann equation \eqref{3.2} is similar to a kinetic equation recently proposed to model the effect of the interstitial fluid on grains in monodisperse gas-solid suspensions \cite{GTSH12}. In this context, our results can be considered of practical interest to analyze transport in a dilute suspension of solid particles (whose collisions are elastic) when the velocity of the particles follows the velocity of the fluid  \cite{TK95}. Moreover, when $\mathbf{U}_\text{g}=\mathbf{0}$ and $\gamma\equiv \text{const.}$, in the case of hard spheres the drag force term $\gamma \partial_\mathbf{v}\cdot \mathbf{v} f$ appearing in Eq.\ \eqref{3.2} arises from a (logarithmic) change in the time scale of the hard sphere system without external force \cite{L01}.

To solve Eq.\ \eqref{3.2}, we consider states that deviate from the time-dependent equilibrium $f_0$ by \emph{small} spatial gradients. In these conditions, the Boltzmann equation \eqref{3.2} can be solved by the Chapman-Enskog method \cite{CC70} conveniently adapted to account for the time dependence of the reference distribution function $f_0(\mathbf{r}, \mathbf{v},t)$. As usual, the Chapman-Enskog method assumes the existence of a \emph{normal} or hydrodynamic solution \cite{CC70} such that all the space and time dependence of the distribution function $f(\mathbf{r}, \mathbf{v},t)$ only occurs through the hydrodynamic fields $n(\mathbf{r},t)$, $\mathbf{U}(\mathbf{r},t)$, and $T(\mathbf{r},t)$:
\beq
\label{3.3}
f(\mathbf{r}, \mathbf{v},t)=f\left[\mathbf{v}|n(\mathbf{r},t), T(\mathbf{r},t), \mathbf{U}(\mathbf{r},t)\right].
\eeq
The notation on the right hand side indicates a functional dependence on density, temperature and flow velocity. In the case of small spatial variations, the functional dependence \eqref{3.3} can be made local in space through an expansion in spatial gradients of the fields. In this case,

%To generate the expansion, $f$ is written as a series expansion in a formal
%parameter $\epsilon $ measuring the nonuniformity of the system, i.e.,
\begin{equation}
f=f^{(0)}+\epsilon \,f^{(1)}+\epsilon^2 \,f^{(2)}+\cdots \;,
\label{3.4}
\end{equation}
where each factor of $\epsilon $ means an implicit gradient of a hydrodynamic field. Moreover, in ordering the different level of approximations in Eq.\ \eqref{3.2}, one has to characterize the magnitude of the drag coefficient $\gamma$ relative to the gradients as well. As in recent previous studies \cite{GTSH12,GCV13,KG13}, we take it to be at least of zeroth-order in gradients. \vicente{Thus, since the friction coefficient can also depend on space, one has to write $\gamma=\gamma^{(0)}+\epsilon \gamma^{(1)}+\cdots$}. A different consideration must be given to the term proportional to the velocity difference $\Delta \mathbf{U}$ in Eq.\ \eqref{3.2} since it is expected that this term contributes to the mass or heat fluxes in sedimentation problems, for instance. In fact, the term $\Delta \mathbf{U}$ can be interpreted as an external field (like gravity) and so, it should be considered at least to be of first order in perturbation expansion.. Finally, according to the expansion \eqref{3.4} for the distribution function, the time derivative and the fluxes are also expanded in powers of $\epsilon$. This is the usual Chapman-Enskog method for solving kinetic equations.

\subsection{Zeroth-order approximation}

To zeroth order in $\epsilon$, the Boltzmann equation \eqref{3.2} for $f^{(0)}$ reads
\beq
\label{3.6}
\partial_{t}^{(0)}f^{(0)} +\gamma^{(0)}
\frac{\partial}{\partial\mathbf{v}}\cdot \mathbf{V}
f^{(0)}= J[f^{(0)},f^{(0)}].
\end{equation}
Upon writing Eq.\ \eqref{3.6} use has been made of the fact that $\Delta \mathbf{U}$ is at least of first order in the gradients.
The balance equations at zeroth order give $\partial_t^{(0)}n=\partial_t^{(0)} U_i=0$ and $\partial_t^{(0)}T=2\gamma^{(0)} T$.
%\begin{equation}
%\label{3.7}
%\partial_t^{(0)}n=0, \quad \partial_t^{(0)}\mathbf{U}=\textbf{0},
%\end{equation}
%\begin{equation}
%\label{3.8}
%\partial_t^{(0)}T=2\gamma T.
%\end{equation}
Using these derivatives, Eq.\ \eqref{3.6} becomes
\beq
\label{3.10}
2\gamma^{(0)} T\frac{\partial f^{(0)}}{\partial T} +\gamma^{(0)}
\frac{\partial}{\partial\mathbf{v}}\cdot \mathbf{V}
f^{(0)}= J[f^{(0)},f^{(0)}].
\eeq
A solution to Eq.\ \eqref{3.10} is given by the local version of the time-dependent Maxwellian distribution $f_0$, namely, $f^{(0)}$ is given by Eq.\ \eqref{2.13} with the replacements $n\to n(\mathbf{r},t)$, $\mathbf{v}\to \mathbf{V}(\mathbf{r},t)$, and $T(t)\to T(\mathbf{r},t)$. The isotropic properties of $f^{(0)}$ yield $P_{ij}^{(0)}=p\delta_{ij}$ and $\mathbf{q}^{(0)}=\mathbf{0}$, where $p=nk_BT$ is the hydrostatic pressure.

\section{Transport coefficients}
\label{sec4}

The analysis to first order in spatial gradients is more involved and follows similar steps as those made in the absence of external forces \cite{CC70}. The kinetic equation for the distribution $f^{(1)}$ is given by
\beqa
\label{4.1}
\left(\partial_{t}^{(0)}+{\cal L}\right)f^{(1)}+\gamma^{(0)}
\frac{\partial}{\partial {\bf v}}\cdot {\bf V}f^{(1)}
&=&-f^{(0)}\left( \frac{mV^2}{2k_B T}-\frac{5}{2}\right) \mathbf{V}\cdot \nabla \ln T \nonumber\\
& & -
f^{(0)}\frac{m}{k_B T}\left( V_i V_j-\frac{1}{3} V^2 \delta_{ij} \right)\nabla_i U_j, \nonumber\\
\eeqa
where ${\cal L}$ is the linearized Boltzmann collision operator
\begin{equation}
\label{4.2}
{\cal L}f^{(1)}=-\left(J[f^{(0)},f^{(1)}]+J[f^{(1)},f^{(0)}]\right).
\end{equation}
\vicente{Note that $\gamma^{(1)}=0$ since the drag coefficient $\gamma$ is a scalar, and corrections to first order in the gradients can arise only from the divergence of the vector field. However, as Eq.\ \eqref{4.1} shows, there is no contribution to $f^{(1)}$ proportional to $\nabla \cdot \mathbf{U}$ and so, the first-order correction to $\gamma$ vanishes}. It must be also remarked that  term $\Delta \mathbf{U}$ does not explicitly appear in the form of the first-order distribution $f^{(1)}$.

The first order contributions to the pressure tensor and the heat flux vector are, respectively
\beq
\label{4.3}
\mathsf{P}^{(1)}=\int\; \dd \mathbf{v}\; m \mathbf{V} \mathbf{V} f^{(1)}(\mathbf{v}), \quad
\mathbf{q}^{(1)}=\int\; \dd \mathbf{v}\; \frac{m}{2} V^2 \mathbf{V} f^{(1)}(\mathbf{v}).
\eeq
The evaluation of these fluxes allows one to identify the transport coefficients. In order to get explicit forms of these coefficients, the dependence of the friction coefficient $\gamma$ on space and time must be chosen. Here, we will assume that  $\gamma(\mathbf{r},t)\propto \nu(\mathbf{r},t)$ where $\nu(\mathbf{r},t)$ is an effective collision frequency of the gas (to be chosen later, see for instance the second identity in Eq.\ \eqref{4.7} for the Boltzmann equation). In particular, for $r^{-\beta}$-repulsive potentials, $\nu \propto n T^q$ where $q=\frac{1}{2}-\frac{2}{\beta}$ \cite{GS03,CC70}. For hard spheres ($\beta \to \infty$), $q=1/2$ while $\nu$ is independent of temperature for Maxwell molecules ($\beta=4$ and so, $q=0$). With this choice, the dimensionless friction coefficient
$\gamma^*\equiv \gamma^{(0)}/\nu$ is constant. Another possible simple choice is $\gamma\equiv \text{const.}$ In this case, $\gamma^*$ depends on space and time through the dependence of $\nu$ on the density and temperature. This was the choice considered in Ref.\ \cite{GSB90} to evaluate the influence of the external force on self-diffusion. The expressions of the transport coefficients when $\gamma\equiv \text{const.}$ are provided in \ref{appB}.

\subsection{Maxwell molecules. Boltzmann equation}

The determination of the transport coefficients requires the evaluation of certain moments of the Boltzmann collision operator.
Unfortunately, these collisional moments can only be \emph{exactly} computed in the particular case of Maxwell molecules, namely,
a repulsive interaction potential of the form $\phi(r)=K/r^4$.
In this case, the collision rate appearing in the Boltzmann collision operator is independent of the relative velocity of the two colliding particles and hence, the collisional moments of
degree $k$ can be expressed in terms of velocity moments of the distribution $f$ of degree equal to or smaller than $k$ \cite{TM80}. This is the great advantage of the Boltzmann equation for Maxwell molecules with respect to other interaction potentials.

Let us determine first the pressure tensor $P_{ij}^{(1)}$. Multiplying both sides of Eq.\ \eqref{4.1} by $m V_i V_j$ and integrating over velocity, one gets
\beq
\label{4.5}
\left(\partial_t^{(0)}+\nu \right)P_{ij}^{(1)}-2\gamma^{(0)} P_{ij}^{(1)}=-p \left(\nabla_i U_j+\nabla_j U_i-\frac{2}{3}\delta_{ij}\nabla \cdot \mathbf{U}\right),
\eeq
where use has been made of the result \cite{GS03,TM80}
\beq
\label{4.7}
\int\; \dd \mathbf{v}\; m V_i V_j\; {\cal L}f^{(1)}=\nu P_{ij}^{(1)}, \quad \nu=3 n A_2.
\eeq
The numerical value of $A_2$ is $A_2\simeq 1.3703 \sqrt{2K/m}$, where $K$ is the constant of the Maxwell interaction potential. The solution to Eq.\ \eqref{4.5} is
\beq
\label{4.8}
P_{ij}^{(1)}=-\eta \left(\nabla_i U_j+\nabla_j U_i-\frac{2}{3}\delta_{ij}\nabla \cdot \mathbf{U}\right),
\eeq
where $\eta$ is the shear viscosity coefficient. The shear viscosity coefficient $\eta$ can be written as $\eta=\eta_0 \eta^*(\gamma^*)$, where $\eta_0=p/\nu$ is the Navier-Stokes shear viscosity of a dilute gas in the absence of the external force ($\gamma=0$). Since $\gamma^*$ does not depend on time for Maxwell molecules, then $\partial_t^{(0)}P_{ij}^{(1)}=2\gamma^{(0)} P_{ij}^{(1)}$ and Eq.\ \eqref{4.5} leads to the simple result $\eta=\eta_0$. Thus, the shear viscosity $\eta$ does not depend on the drag force for Maxwell molecules.

The analysis for the heat flux $\mathbf{q}^{(1)}$ follows similar mathematical steps as those carried out for the pressure tensor. From Eq.\ \eqref{4.1}, one obtains
\beq
\label{4.12}
\left(\partial_t^{(0)}+\frac{2}{3}\nu\right) \mathbf{q}^{(1)}-3\gamma^{(0)} \mathbf{q}^{(1)}=-\frac{5}{2}\frac{nk_B^2 T}{m}\nabla T,
\eeq
where use has been made of the result \cite{GS03}
\beq
\label{4.13}
\int\; \dd \mathbf{v}\; \frac{m}{2}V^2 V_i\; {\cal L}f^{(1)}=\frac{2}{3}\nu \mathbf{q}^{(1)}.
\eeq
Since it is expected that $\mathbf{q}^{(1)}$ contains a term proportional to $\nabla T$, then the action of the operator $\partial_t^{(0)}$ on $\nabla T$ will induce a term proportional to the density gradient due to the dependence of $\gamma^{(0)}$ on $n(\mathbf{r},t)$. Thus, to first order in gradients, the heat flow is given by
\beq
\label{4.14}
\mathbf{q}^{(1)}=-\kappa \nabla T-\mu \nabla n,
\eeq
where $\kappa$ is the thermal conductivity coefficient and $\mu$ is a new coefficient that is absent in the absence of the drag force. As said before, the existence of this new term in the heat flux is only due to the \emph{local} dependence of $\gamma^{(0)}$ through the density. In fact, as shown in \ref{appB}, $\mu=0$ when $\gamma$ is assumed to be constant. A similar contribution to the heat flux proportional to the density gradient is also present in granular gases \cite{BP04,BDKS98,GD99}. In this latter case, the presence of this term is associated with the inelastic character of collisions since the coefficient $\mu$ vanishes for elastic collisions (ordinary gases).

The coefficients $\kappa$ and $\mu$ have the forms $\kappa=\kappa_0 \kappa^*(\gamma^*)$ and $\mu=(\kappa_0 T/n)\mu^*(\gamma^*)$,
where $\kappa_0=(15/4)(nk_B^2T/m \nu)$ is the Navier-Stokes thermal conductivity coefficient in the absence of the external force. Here, $\kappa^*$ and $\mu^*$ are  dimensionless functions of $\gamma^*$ to be consistently determined from Eq.\ \eqref{4.12}.
Since for Maxwell molecules $\kappa^*$ and $\mu^*$ do not depend on temperature, the solution to Eq.\ \eqref{4.12} simply yields
\beq
\label{4.17}
\kappa^*=\frac{1}{1+\frac{3}{2}\gamma^*}, \quad \mu^*=-3\gamma^*\kappa^{*2}.
\eeq
In contrast to the shear viscosity, the thermal conductivity $\kappa$ and the coefficient $\mu$ depend explicitly on the friction coefficient $\gamma$. As expected, $\mu=0$ when $\gamma=0$.

\begin{figure}
%[hbtp]
%\begin{center}
%\begin{tabular}{lr}
%\resizebox{3.7cm}{!}
{\includegraphics[width=0.5\columnwidth]{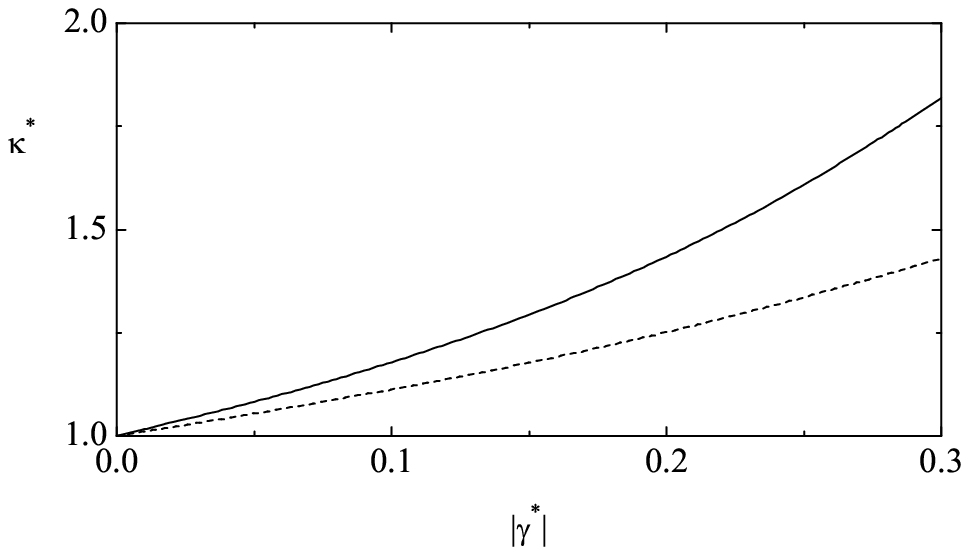}}
%&\resizebox{3.2cm}{!}
{\includegraphics[width=0.49\columnwidth]{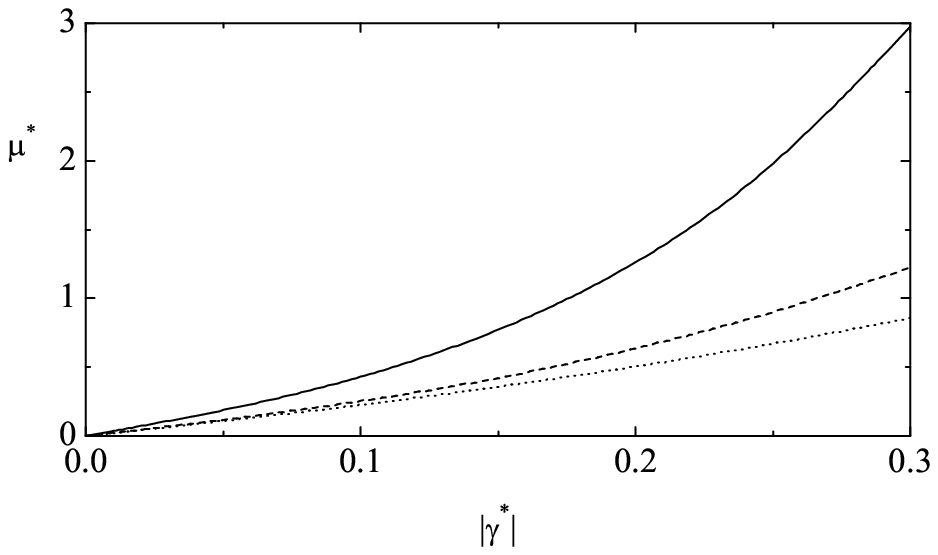}}
%\end{tabular}
%\resizebox{6.5cm}{!}{\includegraphics{fig4.eps}}
%\end{center}
\caption{Plot of the reduced transport coefficients $\kappa^*$ (left panel) and $\mu^*$ (right panel) versus the (dimensionless) friction coefficient $|\gamma^*|$. In the case of $\kappa^*$, the solid line refers to the result obtained from the Boltzmann equation for Maxwell molecules while the dashed line corresponds to the result derived from the BGK kinetic model for general repulsive interaction potentials. In the case of $\mu^*$, the solid line is the result obtained from the Boltzmann equation for Maxwell molecules while the dashed and dotted lines correspond to the results derived from the BGK kinetic model for Maxwell molecules ($q=0$) and hard spheres ($q=\frac{1}{2}$), respectively.
\label{fig1}}
\end{figure}

\subsection{General potentials. BGK kinetic model}
\label{sec4b}

The results derived in the previous Subsection hold for the Boltzmann equation in the particular case of Maxwell molecules. On the other hand, the Chapman-Enskog solution to the Boltzmann equation for interactions different from the Maxwell one is technically difficult. In this case, it is useful to consider kinetic models with the same qualitative features as the Boltzmann equation but with a mathematically simpler structure. The most widely kinetic model used for a dilute gas is the BGK kinetic model \cite{GS03} which is constructed by replacing the true Boltzmann collision operator by a single relaxation-time term:
\beq
\label{4.18.1}
J[f,f]\to -\zeta (f-f_\text{LE}),
\eeq
where $\zeta(\mathbf{r},t)$ is an effective collision frequency (independent of particle velocity) and $f_\text{LE}$ is the local equilibrium distribution function. The collision frequency $\zeta$ can be seen as a free parameter of the model to be chosen to optimize the agreement with the results derived from the Boltzmann equation.

The Chapman-Enskog solution to the BGK kinetic model equation proceeds in the same way as described in the previous subsection. The first-order distribution $f^{(1)}$ is determined from Eq.\ \eqref{4.1} with only the replacement ${\cal L}f^{(1)}\to \zeta f^{(1)}$. With these changes, the corresponding BGK expressions for the transport coefficients are
\beq
\label{4.22}
\eta^*=\frac{1}{1-2q\gamma^*},
\eeq
\beq
\label{4.26}
\kappa^*=\frac{1}{1+\gamma^*}, \quad \mu^*=-\frac{2\gamma^*}{(1+\gamma^*)[1+(1-2q)\gamma^*]}.
\eeq
Equation \eqref{4.26} shows that, in contrast to what happens for $\eta^*$, the BGK predictions for $\kappa^*$ and $\mu^*$ for Maxwell molecules ($q=0$) differ from those obtained from the Boltzmann equation, Eq.\ \eqref{4.17}. \vicente{In addition, since $\gamma^*<0$ then Eqs.\ \eqref{4.22} and \eqref{4.26} yield physical (positive) values of the (reduced) coefficients $\eta^*$, $\kappa^*$, and $\mu^*$ when $|\gamma^*|<1$}.

The dependence of the (reduced) thermal conductivity $\kappa^*\equiv \kappa/\kappa_0$ and the (reduced) coefficient $\mu^*=(\kappa_0 T/n)\mu$ on the (reduced) friction coefficient $|\gamma^*|$ is plotted in Fig.\ \ref{fig1}. The results obtained from the Boltzmann equation for Maxwell molecules are compared with those derived from the BGK model for Maxwell molecules ($q=0$) and hard spheres ($q=\frac{1}{2}$). We observe first that the BGK predictions for Maxwell molecules agree qualitatively well with those obtained from the Boltzmann equation, showing again the reliability of the BGK model kinetic equation. On the other hand, the differences between both predictions increase with the magnitude of the friction coefficient $\gamma^*$, specially in the case of the coefficient $\mu^*$. With respect to the influence of the interaction potential on $\mu^*$, we see that it is in general weak although it becomes more   significant as $|\gamma^*|$ increases.

\section{Stability of the time-dependent equilibrium state}
\label{sec5}

The closed Navier-Stokes hydrodynamic equations for the hydrodynamic fields $n$, ${\bf U}$, and $T$ can be obtained by replacing the constitutive forms of the pressure tensor and the heat flux into the balance equations \eqref{2.5}--\eqref{2.7}. The corresponding Navier-Stokes hydrodynamic equations are analogous to that of an unforced  gas, except for the presence of the term $\gamma^{(0)} \Delta \mathbf{U}$ in the momentum balance equation and the terms $2\gamma^{(0)} T$ and $\mu \nabla n$ in the energy balance equation. In addition, as shown in Section  \ref{sec4}, the shear viscosity $\eta$ and the thermal conductivity $\kappa$ coefficients differ in general from those obtained in the absence of the drag force.

As said in Section \ref{sec2}, the Navier-Stokes hydrodynamic equations admit a simple solution which corresponds to the so-called time-dependent equilibrium state. It describes a uniform state with vanishing flow fields ($\mathbf{U}=\mathbf{U}_g=\mathbf{0}$) and a time-dependent temperature obeying Eq.\ \eqref{2.12}. An interesting question is whether the above time-dependent equilibrium state is stable. In order to perform this analysis, we will assume here that $\mathbf{U}=\mathbf{U}_g$ (and so, the term $\Delta \mathbf{U}=\mathbf{0}$) and consider the realistic case of hard spheres interaction. In this case, we will use the BGK expressions of the transport coefficients for hard spheres ($q=1/2$). The analysis for other interaction potentials follows similar mathematical steps as those made below for hard spheres.

Under the above conditions, we want to carry out a linear stability analysis of the nonlinear Navier-Stokes hydrodynamic equations with respect to the (homogeneous) time-dependent equilibrium state for small initial excitations. We assume that the deviations $\delta y_{\alpha}({\bf r},t)=y_{\alpha}({\bf r},t)-y_{\text{s} \alpha}(t)$ are small, where $\delta y_{\alpha}({\bf r},t)$ denotes the deviations of $n$, $\mathbf{U}$, and $T$ from their values in the homogeneous reference state.
The quantities in this homogeneous state verify $\nabla n_\text{H}=\nabla T_\text{H}=\mathbf{0}$, $\mathbf{U}_\text{H}=\mathbf{0}$, and $\partial_t \ln T_\text{H}=2\gamma^{(0)}$, where the subscript $\text{H}$ means that the quantities are evaluated in the homogeneous state. On the other hand, in contrast to the conventional analysis for unforced gases \cite{RL77}, the linearization of the Navier-Stokes equations around the homogeneous state yields a system of partial differential equations with coefficients that are independent of space but depend on time since the reference state is time-dependent. Hopefully, this time dependence can be eliminated through a change in time and space variables and a scaling of the hydrodynamic fields \cite{BDKS98,G05}.

Let us consider now the following (reduced) time and space variables:
\beq
\label{5.6}
\tau=\int_0^{t}\; \zeta_\text{H}(t') t', \quad \boldsymbol{\ell}=\frac{\zeta_\text{H}(t)}{v_\text{H}(t)}\mathbf{r},
\eeq
where $\zeta_\text{H}\propto n_\text{H} T_\text{H}^{1/2}$ is the BGK collision frequency and $v_\text{H}=\sqrt{2k_B T_\text{H}(t)/m}$ is the thermal velocity. The dimensionless time scale $\tau$ is a measure of the average number of collisions per particle between 0 and $t$ while the dimensionless length scale $\boldsymbol{\ell}$ is proportional to the mean free path of gas particles. A set of Fourier transformed dimensionless variables are then introduced by $\rho_{{\bf k}}(\tau)=
\delta n_{{\bf k}}(\tau)/n_{\text{H}}$, ${\bf w}_{{\bf k}}(\tau)=\delta {\bf U}_{{\bf
k}}(\tau)/\sqrt{T_\text{H}(\tau)/m}$ and $\theta_{{\bf k}}(\tau)=\delta T_{{\bf k}}(\tau)/T_{\text{H}}(\tau)$,
where $\delta y_{{\bf k}\alpha}\equiv \{\rho_{{\bf k}},{\bf w}_{{\bf k}}(\tau), \theta_{{\bf k}}(\tau)\}$ is defined as
\begin{equation}
\label{5.8}
\delta y_{{\bf k}\alpha}(\tau)=\int d \mathbf{r}'\;
e^{-i{\bf k}\cdot \mathbf{r}'}\delta y_{\alpha}
(\mathbf{r}',\tau).
\end{equation}
Note that in Eq.\ \eqref{5.8} the wave vector ${\bf k}$ is dimensionless.

In Fourier space, as happens in the unforced case \cite{RL77}, the transverse velocity components
${\bf w}_{{\bf k}\perp}={\bf w}_{{\bf k}}-({\bf w}_{{\bf k}}\cdot
\widehat{{\bf k}})\widehat{{\bf k}}$ (orthogonal to the wave vector ${\bf k}$)
decouple from the other three modes and hence can be obtained more
easily. Their evolution equation can be written as
\begin{equation}
\label{5.9}
\frac{\partial}{\partial \tau}{\bf w}_{{\bf k}\perp}+\left(\gamma_\text{H}^*+\frac{1}{2} \eta_\text{H}^*k^2\right){\bf w}_{{\bf k}\perp}=0,
\end{equation}
where $\eta_\text{H}^*$ is given by Eq.\ \eqref{4.22} with $q=1/2$. The solution to Eq.\ \eqref{5.9} is
\begin{equation}
\label{5.11}
{\bf w}_{{\bf k}\perp}({\bf k}, \tau)={\bf w}_{{\bf k}\perp}(0)\exp\left(s_\perp(k)\tau\right), \quad
s_\perp(k)=-\left(\gamma_\text{H}^*+\frac{1}{2}\eta_\text{H}^*k^2\right).
\end{equation}
Since $|\gamma_\text{H}^*|<1$ (see Subsection  \ref{sec4b}), then Eq.\ \eqref{5.11} shows that there exists a critical wave number
\beq
\label{5.13}
k_\perp =\sqrt{2|\gamma_\text{H}^*|(1+|\gamma_\text{H}^*|)}
\eeq
that separates two regimes: transversal shear modes with $k>k_\perp$ always decay in time (stable modes) while those with $k<k_\perp$ grow exponentially in time (unstable modes).
\begin{figure}
\includegraphics[width=0.65 \columnwidth,angle=0]{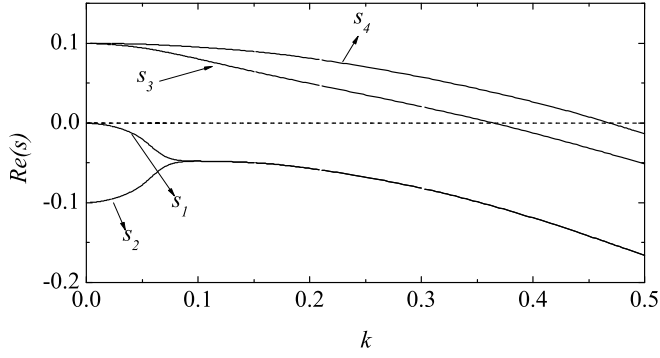}
\caption{Dispersion relations for a dilute gas in the presence of a drag force with $|\gamma_\text{H}^*|=0.1$. From top to bottom the curves correspond to the two degenerate shear (transversal) modes $s_4$ and the remaining three longitudinal modes $s_1$, $s_2$ and $s_3$. Only the real part of the eigenvalues is presented.
\label{fig2}}
\end{figure}

The remaining (longitudinal) modes correspond to $\rho_{{\bf k}}$, $\theta_{{\bf k}}$, and
the longitudinal velocity component of the velocity field, $w_{{\bf k}||}={\bf w}_{{\bf
k}}\cdot \widehat{{\bf k}}$ (parallel to ${\bf k}$). These modes are coupled and obey the equation
\begin{equation}
\frac{\partial \delta y_{{\bf k}\alpha }(\tau )}{\partial \tau }=M_{\alpha \beta}
\delta y_{{\bf k}\beta }(\tau),
\label{5.14}
\end{equation}
where $\delta y_{{\bf k}\alpha }(\tau )$ denotes now the set  $\left\{\rho _{{\bf k}},w_{{\bf k}||},\theta _{{\bf k}}
 \right\}$ and $\mathsf{M}$ is the square matrix
%\begin{widetext}
\begin{equation}
{\sf M}=\left(
\begin{array}{ccc}
0 & -i k& 0 \\
-\frac{1}{2}ik& -\frac{2}{3}\eta_\text{H}^*k^2-\gamma_\text{H}^*& -\frac{1}{2}ik\\
2\gamma_\text{H}^*-\frac{5}{6}\mu_\text{H}^*k^2& -\frac{2}{3} ik & \gamma_\text{H}^*-\frac{5}{6}\kappa_\text{H}^*k^2
\end{array}
\right),   \label{5.15}
\end{equation}
%\end{widetext}
where $\kappa_\text{H}^*$ and $\mu_\text{H}^*$ are given by Eq.\ \eqref{4.26} with $q=1/2$. The longitudinal three modes have the form $\text{exp}[s_\ell (k)\tau]$ for $\ell=1,2,3$, where $s_\ell(k)$ are the eigenvalues of the matrix ${\sf M}$. As in the unforced case \cite{RL77}, it is seen that at very small $k$ all modes are real, while at larger $k$ two modes become a complex conjugate pair of propagating modes. The dispersion relations $s_\ell (k)$ with $|\gamma_\text{H}^*|=0.1$ are plotted in Fig.\ \ref{fig2}. Only the real part (propagating modes) of the eigenvalues is represented. We observe that the heat mode can be also unstable for $k<k_\parallel$, where $k_\parallel$ is obtained from the relation $\det{{\sf M}}=\textbf{0}$. Its expression is
\beq
\label{5.18}
k_\parallel=\sqrt{\frac{6}{5}\frac{|\gamma_\text{H}^*|}{\kappa_\text{H}^*-\mu_\text{H}^*}}.
\eeq

\begin{figure}
\includegraphics[width=0.65 \columnwidth,angle=0]{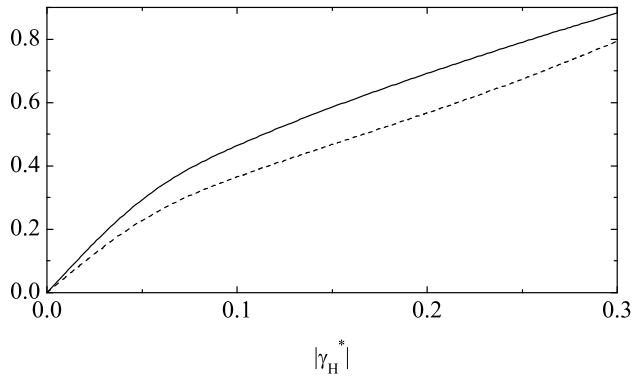}
\caption{Plot of the critical wave numbers $k_\perp$ (solid line) and $k_\parallel$ (dashed line) as functions of the friction coefficient $|\gamma_\text{H}^*|$.
\label{fig3}}
\end{figure}

The dependence of the critical values $k_\perp$ and $k_\parallel$ on $|\gamma_\text{H}^*|$ is illustrated in Fig.\ \ref{fig3}. We observe that the magnitude of both critical values increases with $|\gamma_\text{H}^*|$. Moreover, given that $k_\perp > k_\parallel$, the onset of instability is always driven by the shear mode. Thus, according to the results obtained here, three different regions can be identified. For $k>k_\perp$, all modes are negative and the system is linearly stable with respect to initial perturbations with wave number in this range (short wavelength region). For $k_\parallel<k<k_\perp$, the shear mode is unstable while the heat mode is linearly stable. In this range, the density (which is coupled to the heat mode) is also stable and so, density inhomogeneities can only be generated by the nonlinear coupling with the unstable shear mode. Finally, if $k<k_\parallel$, first vortices and then clusters are developed and the final state of the system is strongly inhomogeneous.

\vicente{It must be recalled that we have scaled the velocity field with the time-dependent thermal velocity $\sqrt{T_\text{H}(t)/m}$ of the homogeneous reference state. Thus, if one considers the (unscaled) velocity $\mathbf{U}$, we have that the perturbation $\delta \mathbf{U}$ decays exponentially in time because of $\partial_t T_\text{H}=2\gamma^{(0)} T_\text{H}$ ($\gamma^{(0)}<0$). This result indicates that the linear analysis will not be sufficient and one has to consider nonlinear terms (such the viscous heating appearing in the balance equation for temperature) accounting for coupling between the different hydrodynamic modes. As happens in granular fluids \cite{BRC99}, it is possible that the above coupling leads to the formation of the clusters in the time-dependent equilibrium state. In any case, the behavior of the transversal component of the scaled velocity field provides the onset of instability since it gives the critical size of the system beyond which the time-dependent equilibrium state becomes unstable}.

\vicente{Before closing this Section, it is important to recall that we have analyzed here the stability of the homogeneous time dependent state rather than the stability of the \emph{steady} states generated in computer simulations with the thermostat. In this latter case, as shown in previous works on driven granular gases \cite{GCV13,Gradenigo11}, the steady reference state turns out to be linearly stable}.

\section{Conclusions}
\label{sec6}

In this paper we studied the effect of a drag force proportional to the particle velocity on the transport coefficients of a dilute gas. As shown in a previous work \cite{GSB90}, in the homogeneous state, when this force acts only on the system then the distribution function of gas reaches in the long time limit a Maxwellian form $f_0(\mathbf{v},t)$ with a time-dependent temperature [see Eqs.\ \eqref{2.12} and \eqref{2.13}]. The transport coefficients of the gas have been obtained by means of a Chapman-Enskog expansion around the local version of the distribution $f_0(\mathbf{v},t)$ instead of the (local) equilibrium distribution function. This is the new feature of the corresponding expansion.

The fact that the temperature is not stationary in the zeroth-order solution (i.e., $\partial_t^{(0)}T \neq 0$) gives rise to conceptual and practical difficulties not present in the absence of the drag force. One of them is that in general the evaluation of the complete nonlinear dependence of the transport coefficients on the friction coefficient $\gamma$ [defined in Eq.\ \eqref{3.1}] requires the numerical integration of the differential equations defining the transport coefficients. This is quite an intricate problem. On the other hand, if one assumes that $\gamma \propto \nu$ (where $\nu$ is an effective collision frequency of the system), then explicit expressions for the above transport coefficients can be obtained. This has been \emph{exactly} accomplished here from the Boltzmann equation for Maxwell molecules while the BGK kinetic model has been used to determine the transport coefficients for non-Maxwell molecules. One of the main differences with respect to the results derived in the conventional case is that the heat flux $\mathbf{q}$ does not obey Fourier's law since there is an additional contribution to $\mathbf{q}$ proportional to the density gradient. The existence of this new term in the heat flux is also present in the case of granular gases \cite{G03,CW07}. Thus, our model (model of elastic spheres subject to a drag force) can be seen as an alternative \cite{AA05} to the well-known inelastic hard sphere model for granular media.

The knowledge of the transport coefficients allows one to solve the closed set of hydrodynamic equations for situations close to
the time-equilibrium state described by the distribution $f_0(\mathbf{v},t)$. A linear stability analysis of those hydrodynamic equations have been carried out and the corresponding conditions for instability have been identified in terms of the friction coefficient $\gamma$. Our results show that the reference state $f_0$ can be unstable and this instability is driven by the transversal shear mode (see Fig.\ \ref{fig3}). It must be noted that the above instability is absent when the external force $\mathbf{F}$ is proportional to the particle velocity $\mathbf{v}$ instead of the peculiar velocity $\mathbf{V}$. This result is consistent with the findings of Ref.\ \cite{GSB90} where an $H$-theorem was proved for the distribution $f_0(t)$ when $\mathbf{F} \propto \mathbf{v}$.

A possible direction of study is to extend the analysis made here for ordinary gases to the important subject of granular gases. As a first step, one could consider the so-called inelastic Maxwell models whose tractability opens the possibility of obtaining exact results in some problems. Previous works carried out by one of the authors and co-workers \cite{IMM} have clearly shown the reliability of inelastic Maxwell models to reproduce some trends observed in real granular flows.

%\vspace{0.1cm}

\begin{center}
\textbf{Acknowledgments}
\end{center}
The research of V.G. has been supported by the Ministerio de
Educaci\'on y Ciencia (Spain) through grant No. FIS2010-16587, partially financed by
FEDER funds and by the Junta de Extremadura (Spain) through Grant No. GRU10158.

\appendix
\section{Transport coefficients when $\gamma$ is constant}
\label{appB}

In this Appendix we provide the explicit expressions of the transport coefficients when the friction coefficient $\gamma$ is constant. In this case, in contrast to the results derived in Section \ref{sec4}, $\gamma^* \propto \nu^{-1}\propto n^{-1} T^{-q}$ so that $T\partial_T \gamma^* \propto -q \gamma^*$. The expressions for the dimensionless functions $\eta^*$, $\kappa^*$, and $\mu^*$ can be easily determined for this choice. In particular, $\mu^*=0$ for any interaction potential and so, the conventional Fourier's law is recovered.

In the context of the Boltzmann equation for Maxwell molecules, the expressions of $\eta^*$ and $\kappa^*$ are the same as those obtained in Section \ref{sec4}, namely, $\eta^*=1$ and $\kappa^*$ is given by Eq.\ \eqref{4.17}. On the other hand, the BGK results differ from those derived when $\gamma \propto \nu$ since $\eta^*$ and $\kappa^*$ obey the following nonlinear differential equations:
\beq
\label{b2}
2\gamma^*\left(1-q\gamma^*\frac{\partial}{\partial \gamma^*}\right)\eta^*+(1-2\gamma^*)\eta^*=1,
\eeq
\beq
\label{b3}
2(1-q)\gamma^*\kappa^*-2q\gamma^{*2}\frac{\partial \kappa^*}{\partial \gamma^*}+(1-\gamma^*)\kappa^*=1,
\eeq
where the particular case of $r^{-\beta}$ potentials has been considered. It is quite apparent that Eqs.\ \eqref{b2} and \eqref{b3} must be solved in general numerically to obtain the dependence of $\eta^*$ and $\kappa^*$ on $\gamma^*$. An exception corresponds to Maxwell molecules ($q=0$), in which case one simply gets $\eta_{\text{M}}^*=1$ and $\kappa_{\text{M}}^*=(1+\gamma^*)^{-1}$. For $q\neq 0$, based on previous results derived for Maxwell molecules and hard spheres in the absence of external nonconservative forces\cite{K97,S00}, it is expected that there is a weak influence of the interaction potential on transport properties. This suggests to expand the transport properties in powers of the interaction parameter $q$ as an alternative to obtain accurate analytical results for non-Maxwell molecules. Thus, for non-Maxwell molecules, we consider the simplest approximations $\eta^*\simeq \eta_{\text{M}}^*+\eta_1 q$ and $\kappa^*\simeq \kappa_{\text{M}}^*+\kappa_1 q$. Inserting the above expansions into Eqs.\ \eqref{b2} and \eqref{b3}, respectively, and neglecting nonlinear terms in $q$ one obtains $\eta_1=0$ and $\kappa_1=2\gamma^*/(1+\gamma^*)^3$.

%%%%%%%%%%%%%%%%%%%%%%%%%%%%%%%%%%%%%%%%%%%%%%%%%%%%%%%%%%%%%%%%%%%%%%%%%%%%%%%%%%%%%%%%%%%%%%%%%%%%%%%%%%%%%%%%%%%%%%%%%%%%%%%%%%%%

\end{document}